\documentclass[aps,prl,superscriptaddress,showpacs,floatfix,preprint]{revtex4}

\usepackage{amsmath,graphicx,color,amssymb}

\newcommand{\bl}[1]{\begin{equation}\label{#1}}
\newcommand{\ee}{\end{equation}}
\newcommand{\bea}{\begin{eqnarray}}
\newcommand{\eea}{\end{eqnarray}}
\newcommand{\pd}[2]{\frac{\partial{#1}}{\partial{#2}}}
\newcommand{\rec}[1]{\frac{1}{#1}}
\newcommand{\td}[2]{\frac{\mathrm{d}{#1}}{\mathrm{d}{#2}}}
\newcommand{\z}[1]{\left({#1}\right)}
\newcommand{\sz}[1]{\left[{#1}\right]}
\renewcommand{\v}[1]{\mathbf{#1}}

\renewcommand{\r}[1]{(\ref{#1})}

\newcommand{\elte}{ELTE, E{\"o}tv{\"o}s Lor{\'a}nd University, H - 1117 Budapest, P{\'a}zm{\'a}ny P. s. 1/A, Hungary}
\newcommand{\kfki}{MTA KFKI RMKI, H-1525 Budapest 114, POBox 49, Hungary}
\newcommand{\stonych}{Department of Chemistry, SUNY Stony Brook, Stony
Brook, NY, 11794-3400, USA}

\begin{document}

\title{New Family of Simple Solutions of
Relativistic Perfect Fluid Hydrodynamics~\footnote{Dedicated to Y. Hama on the occasion of his 70th birthday.}}

\author{T.~Cs{\"o}rg\H{o}}      \affiliation{\kfki}
\author{M.~I.~Nagy}        \affiliation{\elte}
\author{M.~Csan\'ad}    \affiliation{\elte}\affiliation{\stonych}

\begin{abstract}
A new class of accelerating, exact and  explicit solutions of
relativistic hydrodynamics is found --- more than 50 years after the
previous similar result, the  Landau-Khalatnikov solution.
Surprisingly,  the new solutions have a simple form, that
generalizes the renowned, but  accelerationless, Hwa-Bjorken
solution. These new solutions take into account the work done by the
fluid elements on each other, and work  not only in one temporal and
one spatial dimensions, but also in arbitrary number of spatial
dimensions. They are applied here for an advanced estimation of
initial energy density and life-time of the reaction in
ultra-relativistic heavy ion collisions.
\end{abstract}

\pacs{24.10.Nz,47.15.Hg}
\maketitle
\date{\today}

Fluid dynamics is beautifully simple: it is based only on
local conservation of charge, momentum and  energy as well as on
the additional assumption of local thermal equilibrium.
Consequently, the hydrodynamical equations do not have internal scale,
and their applications range from the smallest experimentally accessible
scales of physics, such as the perfect fluid, a new form of matter created
in high energy heavy ion collisions at the Relativistic Heavy
Ion Collider (RHIC) located in Brookhaven National Laboratory (BNL, USA),
through a glass of wine and through astrophysical and stellar objects,
like stellar nebulae, to the largest known object, the evolution of our Universe.
Fluid dynamics is beautifully and sometimes horrendously complicated:
non-linear terms lead to instability, chaos, complexity,
formation of eddies and other beautiful flow patterns,
that are observable from the smallest to the largest scales of physics,
like elliptic flow in high energy heavy ion collisions,
hurricanes, super-nova explosion, or, the Hubble flow of our Universe.
Fluid dynamics is beautifully relevant, too: presently this theory yields the best
description of the single particle momentum distributions,
elliptic flow patterns, and two-particle correlation data
of the thousands of elementary particles created in the Little Bangs of
relativistic heavy ion collisions at RHIC.

It is a matter of fact, that all but one  of the presently known exact
relativistic hydro solutions lack an important feature: the
acceleration of matter.
The only exception is the famous Landau-Khalatnikov  (LK) solution,
discovered more than 50 years ago~\cite{Landau:1953gs,Khalatnikov,Belenkij:1956cd}.
This solution is a 1+1 dimensional, implicitly formulated
but fully analytic solution of relativistic hydrodynamics.
It predicts a realistic, approximately Gaussian
rapidity distribution. However, due to its extremely complicated nature,
the LK solution does not allow for an estimation of the initial energy density.
Another renowned  and exact solution is the 1+1 dimensional,
 boost-invariant, accelerationless
Hwa-Bjorken (HB) solution~\cite{Hwa:1974gn,Bjorken:1982qr}. This
solution allowed Bjorken to give a simple estimate of the initial
energy density reached in heavy ion collisions from final state
hadronic observables. It is well known, that the HB solution (in
its original form, for $\mu_B=0$) leads to
 a flat rapidity distribution, which is at variance with present observations at RHIC,
except perhaps when observations are limited to a narrow region around mid-rapidity.
Acceleration effects are,
however, important in the estimation of the initial energy density
even at mid-rapidity, if the expanding system is finite: even the most central
fluid element performs work on the volume elements closer to the surface,
and this work reduces the internal energy of cells even at mid-rapidity.
We present such a new, accelerating family of solutions below,
and apply it to data analysis in Au+Au collisions at RHIC.
It can also be applied to test
numerical solutions of relativistic hydrodynamics
- no finite, accelerating, exact solution was available before
for such tests in 1+3 dimensions.

\paragraph{Notation and basic equations:}
The metric tensor is $g^{\mu\nu} = \mbox{\it diag}(1,-1,-1,-1)$, $u^{\mu}=\gamma(1,\v{v})$ is the
four-velocity field, $\v{v}=v\v{n}$ is the three-velocity. The pressure is denoted
by  $p$\,, the energy density by $\varepsilon$\,, the
temperature by  $T$\,, the charged particle density by $n$\,, the chemical potential
by $\mu$ and the entropy density by $\sigma$.

In high energy collisions, the entropy density is large, but
net charge density is small.
In perfect fluids, entropy and four-momentum are locally  conserved,
\bea
\partial_{\nu} (\sigma u^{\nu} )  & =&  0 , \label{e:cont} \\
\partial_{\nu}T^{\mu\nu} & = &  0 ,  \label{EMT}
\eea
where the energy-momentum tensor is
\bl{EMT-0}
T^{\mu\nu}=(\varepsilon + p) u^{\mu}u^{\nu}-pg^{\mu\nu}.
\ee
The relativistic Euler equation and the energy conservation
law are projections of Eq.~(\ref{EMT}):
\bea
(\varepsilon+p)u^{\nu}\partial_{\nu}u^{\mu}&=& \z{g^{\mu\rho}-u^{\mu}u^{\rho}}\partial_{\rho}p, \label{Reul} \\
(\varepsilon + p)\partial_{\nu}u^{\nu}+u^{\nu}\partial_{\nu}\varepsilon &=& 0. \label{RE}
\eea
For simplicity, let us consider the case, when all the conserved charges $c_i$ have 
$\mu_i = 0$.
Such an approximation is common in
high energy physics at RHIC, and is assumed both in the LK
and in the HB solutions.
The thermodynamics of the flowing matter is
characterized by the Equations of State (EoS).
Let us consider
\bl{EOS}
\varepsilon - B = \kappa (p + B),
\ee
where $B$ stands for the bag constant and $\kappa = 1/c_s^2$, where $c_s$ stands for the speed of sound,
$ c_s^2= dp/d\varepsilon$. The bag constant $B$ may have either a vanishing or a  non-vanishing value, characteristic for
a hadronic or, for a pre-hadronic state,  respectively.  In what follows, $d$ stands for the number of spatial dimensions.
In a heavy ion collision, $d= 3$, irrespective of the characteristics of the flow pattern.
The $\kappa=d$ EoS corresponds
to a gas of massless particles (e.g photons) or an ultra-relativistic ideal gas of
massive particles. In this case, $\sigma \propto T^{d}$ is also true. 
Observe, that Eq.~(\ref{EOS}) closes
Eqs.~(\ref{Reul},\ref{RE}) for the pressure and the three independent
components of velocity.

We discuss below exact solutions of relativistic perfect fluid hydrodynamics in 1+1 dimensions  and
spherical solutions in 1+$d$ dimensions as well.
The notation $r$ stands for the $r_z$
spatial coordinate in 1+1 dimensions,
and for the radial coordinate in 1+$d$ dimensions.
We use the well-known Rindler coordinates ($\tau$ and $\eta$), which naturally fit
to the Hwa-Bjorken solution in the forward light-cone:
\bl{Rindler}
r= \tau\sinh\eta \quad , \quad t = \tau\cosh\eta .
\ee
We rewrite the equations of hydrodynamics in Rindler coordinates for a special case,
when $v=\tanh\Omega(\eta)$, i.e. the $\Omega$ fluid rapidity depends only on $\eta$.
Eqs.~(\ref{Reul}-\ref{EOS}) the yield the following   equations for
$p(\tau,\eta)$ and $\Omega(\eta)$:
\bea
(\kappa+1)\td{\Omega}{\eta} & = &-\frac{\tau}{p}\pd{p}{\tau}-\coth(\Omega-\eta)\rec{p}\pd{p}{\eta} ,
\label{e:EulR}\\
\frac{\kappa+1}{\kappa}\td{\Omega}{\eta} & = & -\frac{\tau}{p}\pd{p}{\tau}-
\tanh(\Omega-\eta)\rec{p}\pd{p}{\eta}- \frac{\kappa+1}{\kappa}
\frac{d-1}{\sinh\eta}\frac{\sinh\Omega}{\cosh(\Omega-\eta)} . \label{ER}
\eea

In what follows, we specify full solutions of relativistic hydrodynamics,
that are valid in the complete forward light-cone. 

\begin{figure}
\includegraphics[height=240pt,angle=-90]{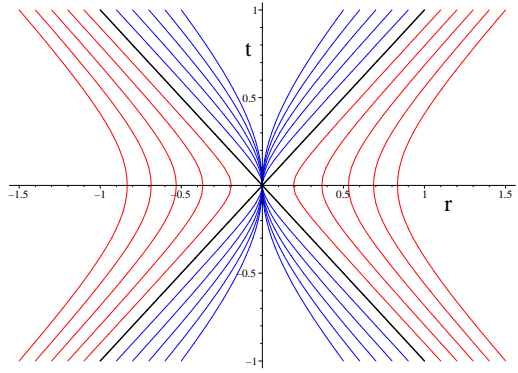}
\caption{\label{fig:0} (Color online) Fluid trajectories of the $\lambda = 2$ solution.
}
\end{figure}

\paragraph{The new class of accelerating solutions} is given by the
following velocity and pressure fields:
\bea
v&=&\tanh\,\lambda\eta , \label{e:veta}\\
p&=&p_0\z{\frac{\tau_0}{\tau}}^{\lambda
d\frac{\kappa+1}{\kappa}}\z{\cosh\frac{\eta}{2}}^{-(d-1)\phi_{\lambda}} - B . \label{e:p}
\eea
The constants $\lambda$, $d$, $\kappa$ and
$\phi_{\lambda}$ are constrained, lines (a)--(e) of Table~\ref{t:1} show the cases
that satisfy the hydrodynamical equations.  Line (a) stands for the well known Hwa-Bjorken
solution~\cite{Hwa:1974gn,Bjorken:1982qr}, also called as Hubble solution when $d = 3$. 
Our new, $\lambda\neq 1$ solutions are listed in lines (b)--(e) of Table~\ref{t:1}.
They describe accelerating flow, indicated by the curvature of the fluid world lines
and expressed mathematically by $u^\mu \partial_\mu u^\nu \neq 0$.
Case (b), where $\lambda=2$  is shown in Fig.~\ref{fig:0}.
Its fluid world lines, $r(t)$ evolve as
\bl{newtrajec}
r(t)=\rec{a_0}(\sqrt{1+(a_0 t)^2} + 1) \quad , \quad a_0=\frac{2 r_0}{\left|r_0^2-t_0^2 \right|} ,
\ee
where $r_0$ and $t_0$ specify the initial condition. 
These are trajectories with constant $a_0$ acceleration in the local rest frame.
Both the $\lambda=1$\,, and  $\lambda=2$ cases can be extended to 
external ($|{\bf r}| > t$) solutions that 
are uniformly accelerating, hence they contain event horizons.
This property of the uniform acceleration was utilized recently by
Kharzeev and Tuchin~\cite{Kharzeev:2005iz} to describe thermalization
in heavy ion reactions via the Unruh effect.
Case (c) describes a one dimensional fluid,  with a special EoS of $\kappa = 1$,
but its parameter of acceleration, $\lambda $ can be chosen arbitrarily.
After our derivation of cases (b) and (c),
T. S. Bir\'o pointed out~\cite{Biro:2007pr}, that the $\lambda=1/2$, $\kappa=1$
and the $\lambda=3/2$, $\kappa=11/3$, $d= 3$ cases are also solutions.
We have generalized them for any $d\in\mathbb{R}$, as shown in lines (d) and (e). 
The exponent $\phi_\lambda$ is introduced to indicate, that the pressure is
explicitly dependent on space-time rapidity $\eta$ in these
cases, and  the pressure tends to
zero for large values of $ |\eta | $. Thus cases (d) and (e)  
are finite solutions. Note that in $d=3$ dimensions, case (e) has
$\varepsilon - 3 p  > 0$,  similarly to   the 
lattice QCD EoS.

Even when one considers the case, that $\lambda$ can be a $\lambda(\tau)$ function, 
and the pressure may have a form of $p=H(\tau)U(\eta) -B$, 
that conditions generalize  the actual solutions of eqs.~(\ref{e:veta},\ref{e:p}),
we have proven  that no non-trivial, additional to Table~\ref{t:1} solutions exist
in this more general class.
Our proof was easily obtained from a second order Taylor 
expansion of the hydrodynamical equations, but is not detailed here.

\begin{table}
\begin{tabular}{|c|c|c|c|c|}
  \hline
  \hline
 Case: & $\lambda$ & $d$ & $\kappa$ & $\phi_{\lambda}$ \\
  \hline
  \hline
 (a)  & $1$             & $\in\mathbb{R}$ & $\in\mathbb{R}$  & $0$ \\ \hline
 (b)  & $2$             & $\in\mathbb{R}$ & $d$              & $0$ \\ \hline
 (c)  & $\in\mathbb{R}$ & $1$             & $1$              & $0$ \\ \hline
 (d)  & $\rec{2}$       & $\in\mathbb{R}$ & $1$              & $\frac{\kappa + 1}{\kappa}$ \\ \hline
 (e)  & $\frac{3}{2}$   & $\in\mathbb{R}$ & $\frac{4d-1}{3}$ & $\frac{\kappa + 1}{\kappa}$ \\ 
  \hline
  \hline
\end{tabular}
  \caption{The new family of solutions is given by lines (b)-(e), while 
case (a) is the Hwa-Bjorken-Hubble solution.}\label{t:1}
\end{table}
During the time evolution of
heavy ion collisions at RHIC, and during the quark-hadron transition in the early Universe,
the chemical potentials of conserved charges are very small,
$p \simeq p(T, \mu_i = 0)$.
In this case, $(1+\kappa)p=\sigma T$. The entropy conservation, Eq.~(\ref{e:cont})
can be solved for $\sigma $, and comparing with $(1+\kappa)p=\sigma T$ yields
\bl{solcont}
\sigma=\sigma_0 \nu_\sigma(s) \z{\frac{p+B}{p_0}}^{\frac{\kappa}{\kappa+1}} ,
    \quad
T= \frac{T_0}{\nu_\sigma(s)}  \z{\frac{p+B}{p_0}}^{\frac{1}{\kappa+1}} ,
\ee
where $(1+\kappa)p_0=T_0\sigma_0$, and $p(\tau,\eta)$ is given by Eq.~\r{e:p}. Here we have introduced
a scaling function $\nu_\sigma(s)$, that can be any positive function that
satisfies $\nu_\sigma(0) = 1$. The scaling variable $s$
has, by definition, a vanishing comoving derivative: $\pd{s}{t}+(\v{v}\nabla)s=0$.
For $\lambda=1$ we find that $s(\tau,\eta)=\eta$. For $\lambda\neq 1$ the scaling variable is
\bl{sdef}
s(\tau,\eta)=\z{\frac{\tau_0}{\tau}}^{\lambda-1}\sinh\z{\z{\lambda-1}\eta} .
\ee
 The scaling function $\nu_\sigma(s)>0$ appears similarly
how it shows up also in accelerationless  solutions~\cite{Csorgo:2003rt,Csorgo:2003ry,Biro:2000nj,Sinyukov:2004am}.

{\it In case of mixtures}\,, where various non-vanishing conserved charges $n_i$ with $\mu_i\neq 0$
are present and contribute to the pressure,
the more general form of the thermodynamic potential,
$p = p(T, \mu_i) = (T \sigma + \sum_i \mu_i n_i)/(1+\kappa) - B $
leads to similar exact solutions of relativistic hydrodynamics,
but new, arbitrary scaling functions $\nu_i(s)>0$
appear, with $n_i\propto \nu_i(s)$ and $\mu_i\propto \rec{\nu_i(s)}$.
These forms solve the continuity equations
for $n_i$ and Eqs.~(\ref{Reul}-\ref{RE}).

\begin{figure}
\includegraphics[width=200pt]{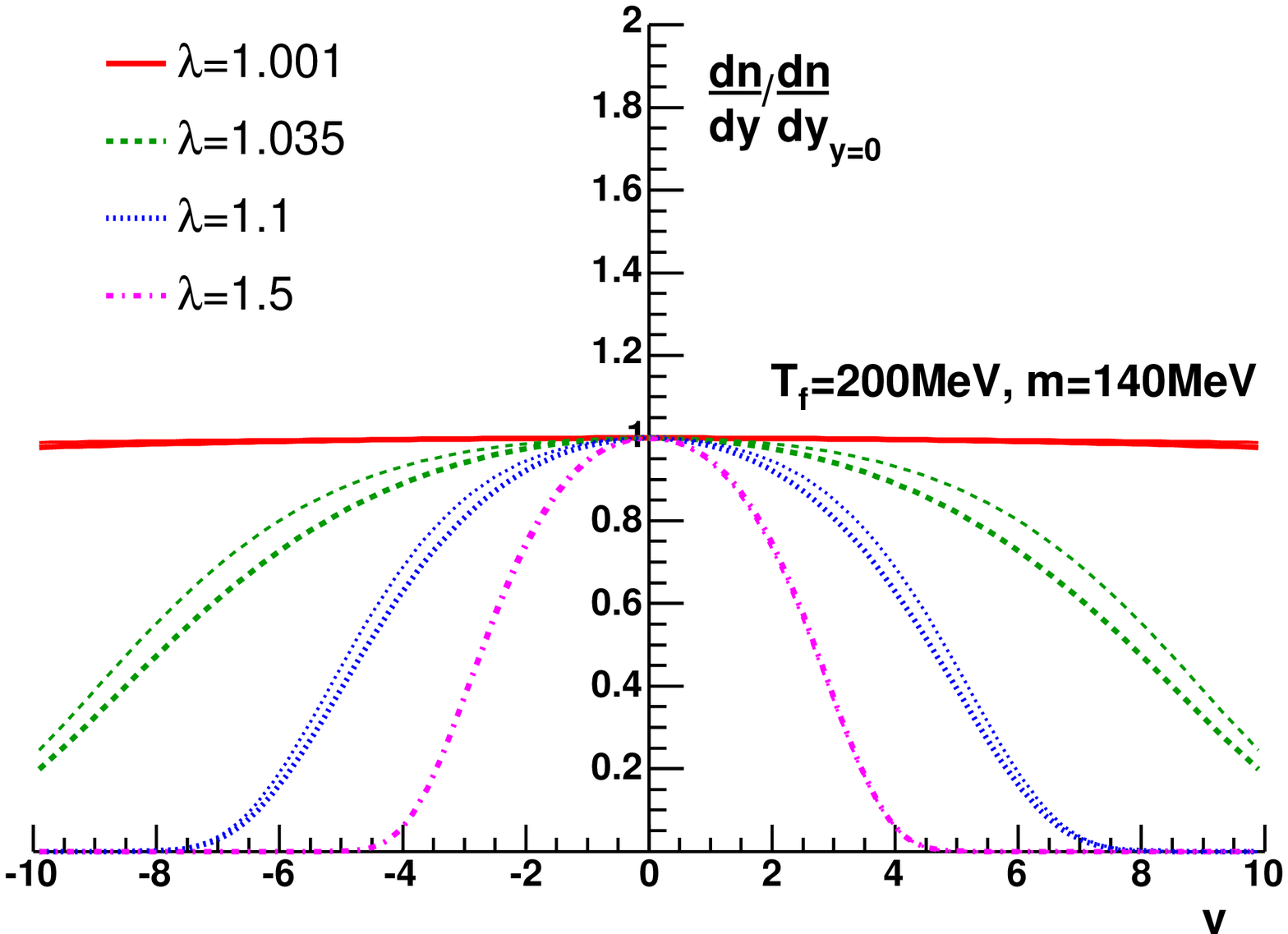}
\includegraphics[width=200pt]{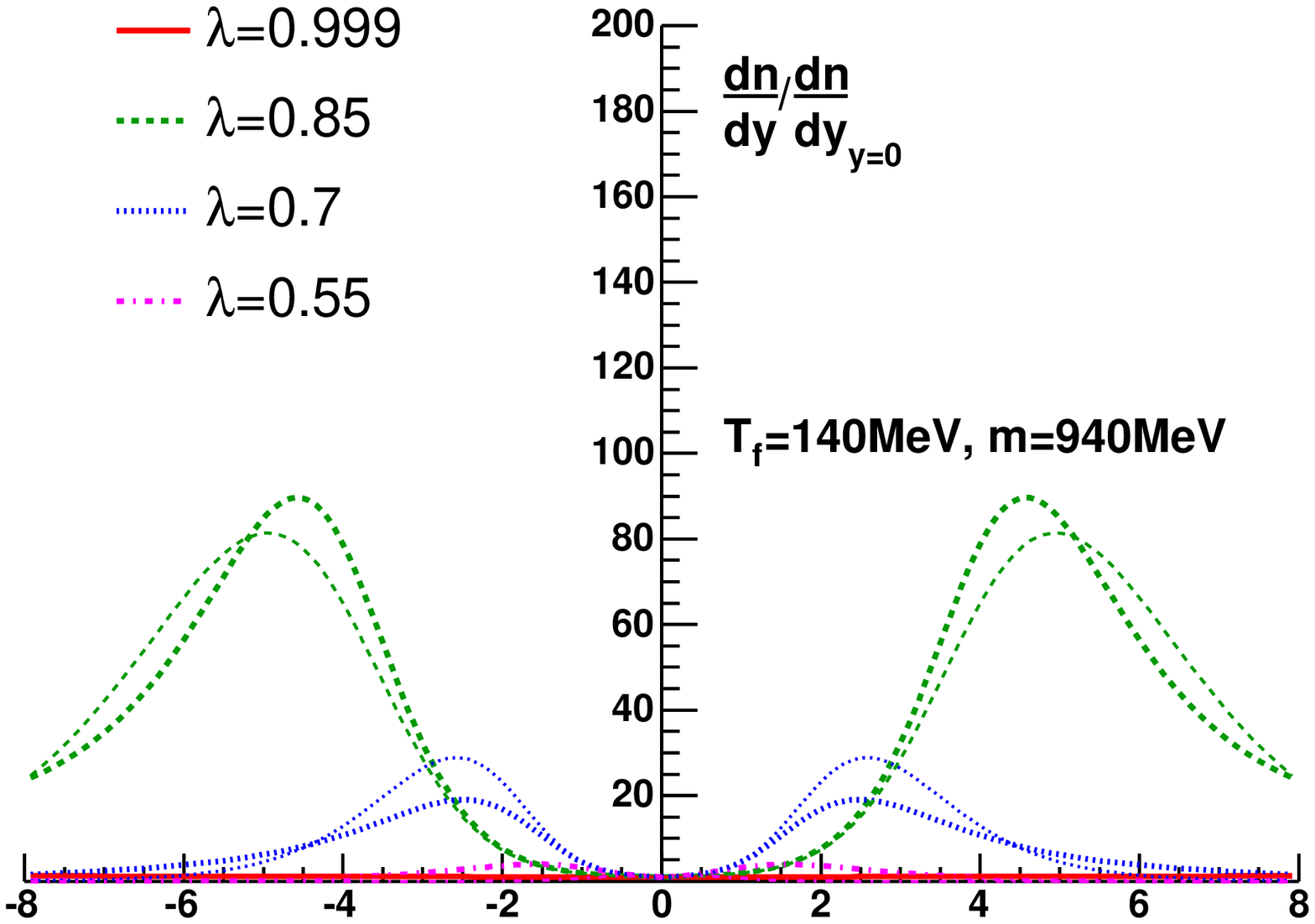}
\caption{\label{fig:1}(Color online) Normalized rapidity
distributions from the new solutions in 1+1 dimensions for various
$\lambda$, $T_f$ and $m$ values. Thick lines
show the result of numerical integration, thin lines
the analytic approximation from Eq.~\r{e:dndy-approx}. For $\lambda>1$
and not too big $T_f$ it can be used within about 10 \% error. }
\end{figure}

\emph{The rapidity distribution}, $\td{n}{y}$ is given below for case (c) of 
Table~\ref{t:1} in a  Boltzmann approximation. We consider the $\nu_\sigma(s) = 1$
case, when our solutions also  solve the Landau-Khalatnikov
equation, $\partial^\nu T u_\mu = \partial_\mu T u^\nu$. 
The freeze-out temperature is  $T(\eta=0,\tau=\tau_f ) = T_f$, 
where subscript $_f$ stands for {\it f}reeze-out.
We assume, that
the freeze-out hypersurface is pseudo-orthogonal to $u^{\mu}$.
With a saddle-point integration in $\eta$,
 for $m/T_f \gg 1$, where $m$ is the particle mass,
$\lambda > 0.5$, $\mu_i=0$  and $\nu_{\sigma}(s)=1$
we got
\bl{e:dndy-approx}
\td{n}{y}\approx
\td{n}{y}\Big{|}_{y=0}
    \cosh^{\pm\frac{\alpha}{2}-1}\z{\frac{y}{\alpha}}
    e^{-\frac{m}{T_f}\sz{\cosh^\alpha\z{\frac{y}{\alpha}} -1 } } ,
\ee
with $\alpha=\frac{2\lambda-1}{\lambda-1}$. 
The ``Gaussian width'' of this distribution is
$
    \Delta y^2 = \frac{\alpha }{m/T_f \mp 1/2 + 1/\alpha} .
$
The upper sign is for the $1+1$ dimensional case, the lower sign is for the case when the $1+1$
dimensional solution is embedded in the 1 + 3 dimensional space-time.
In this latter case, the transverse mass distribution is integrated in a saddle-point approximation from $m$ to
infinity. 
The resulting  rapidity distribution has a  minimum at $y=0$, if $\Delta y^2<0$, it is flat if $\Delta y^2=0$, i.e. $\lambda = 1$,
or $\lambda = \rec{2}\z{1+\frac{T_f}{2m\mp T_f+T_f}}$, otherwise it 
is nearly Gaussian, as illustrated in Fig.~\ref{fig:1}.

\begin{figure}
\includegraphics[width=200pt]{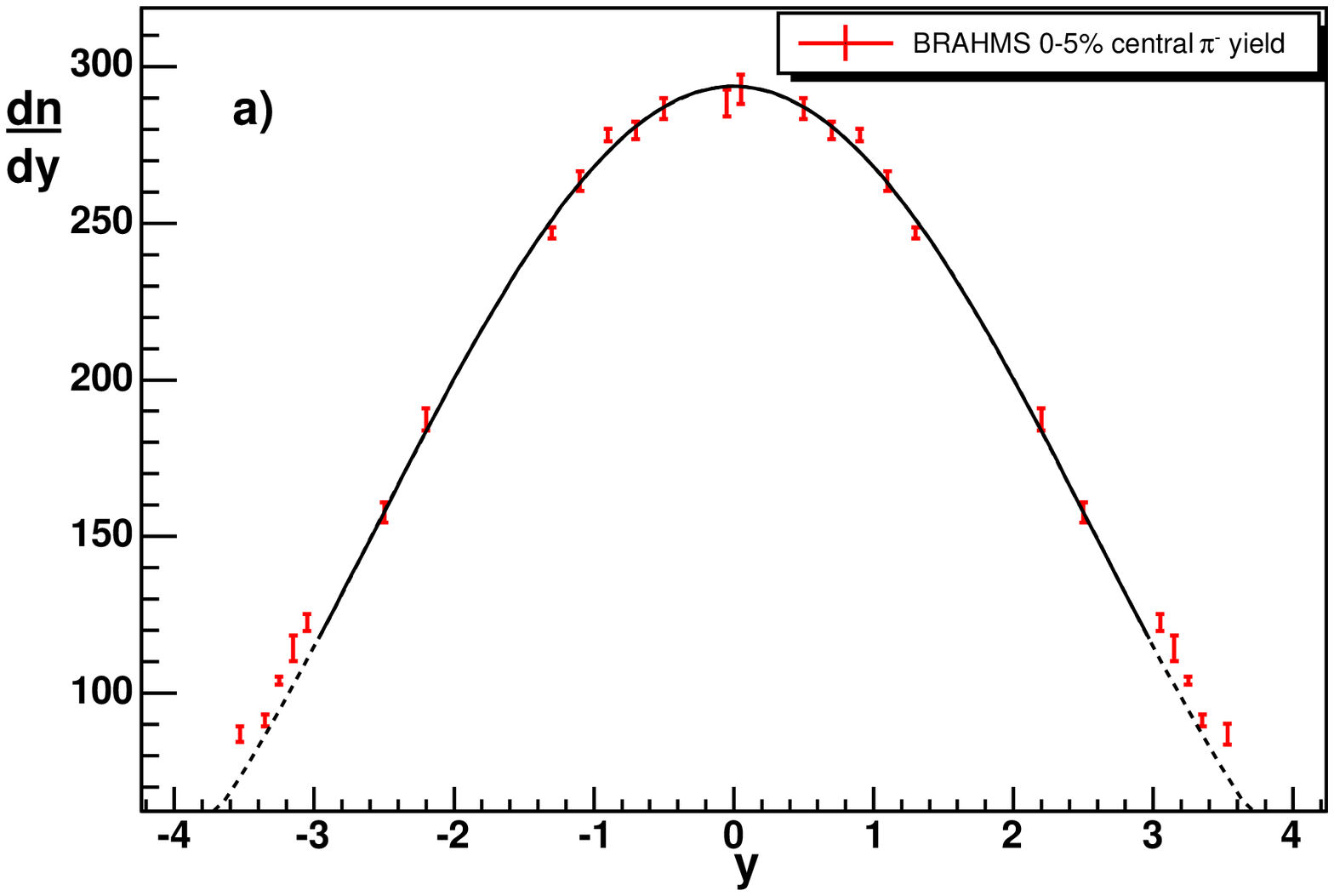}
\includegraphics[width=200pt]{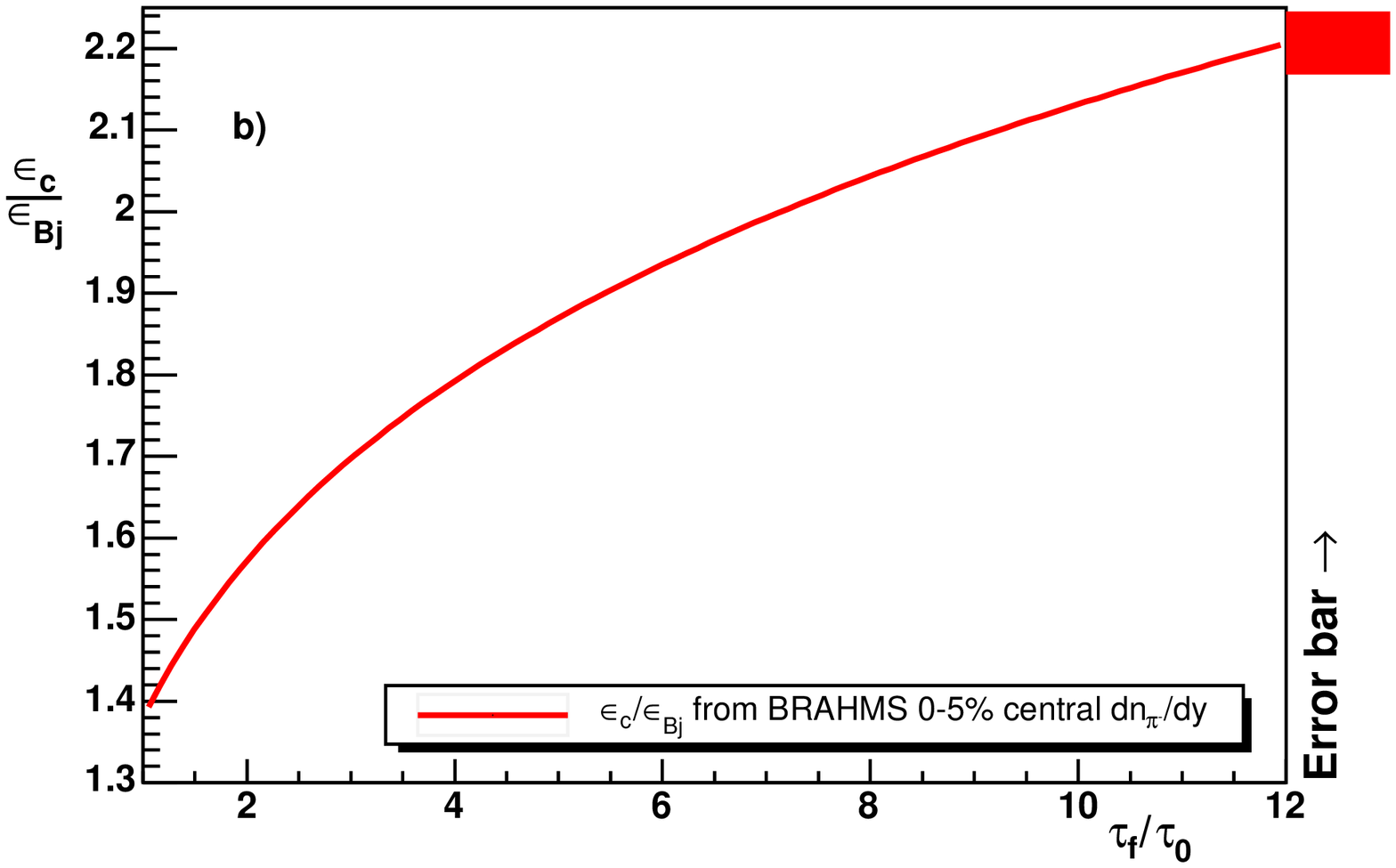}
\caption{\label{fig:3}(Color online)
Panel (a):
$dn/dy$ data of negative pions, as measured by
the BRAHMS collaboration~\cite{Bearden:2004yx} in central (0-5\%)
Au+Au collisions at $\sqrt{s_{NN}}=200$ GeV, fitted with Eq.~\r{e:dndy-approx} (1+3 dimensional case).
The fit range was $-3 < y < 3$, to exclude
target and projectile rapidity region, CL = 0.6 \%.
Panel (b): $\varepsilon_c/\varepsilon_{Bj}$ ratio as a function
of $\tau_f/\tau_0$.
}
\end{figure}

Let us now estimate the energy density reached in heavy ion reactions,
just after thermalization ($\tau=\tau_0 \approx 1$ fm/c).
Let us focus on a thin transverse piece of produced matter at mid-rapidity,
illustrated by Fig. 2 of Ref.~\cite{Bjorken:1982qr}. The radius $R$ of this slab is estimated by the
radius of the colliding hadrons or nuclei, its volume is $dV=(R^2\pi)\tau d\eta$.
The energy content in this slab is $dE = \langle m_t\rangle dn$,
where $\langle m_t \rangle$ is the average
transverse mass at $y=0$, so similarly to Bjorken, the initial energy density is
\bl{e:Bjorken}
    \varepsilon_0 = \frac{\langle m_t\rangle}{ (R^2 \pi)  \tau_0 }
    \frac{dn}{d\eta_0} .
\ee For accelerationless, boost-invariant Hwa-Bjorken flows
$\eta_0=\eta_f=y$, however, for our accelerating solution we have
to apply a correction factor of $\frac{\partial \eta_f}{\partial
\eta_0}\,\frac{\partial y}{\partial
\eta_f}=\z{\tau_f/\tau_0}^{\lambda -1}\z{2\lambda-1}$. Thus the
initial energy density $\varepsilon_0$ can be accessed by an
advanced estimation $\varepsilon_c$ as \bl{e:ncscs}
\frac{\varepsilon_c}{\varepsilon_{Bj}}=\z{2\lambda-1}\z{\frac{\tau_f}{\tau_0}}^{\lambda-1}\,,\,\quad
\varepsilon_{Bj}=\frac{\langle
m_t\rangle}{(R^2\pi)\tau_0}\frac{dn}{dy} . \ee Here
$\varepsilon_{Bj}$ is the Bjorken estimation, which is recovered
if $\td{n}{y}$ is flat (i.e. $\lambda=1$), but if $\lambda>1$,
$\varepsilon_0$ is \emph{under-estimated} by the Bjorken formula.
Fig.~\ref{fig:3} shows our fits to BRAHMS $dn/dy$
data~\cite{Bearden:2004yx}. Using the Bjorken estimate of
$\varepsilon_{Bj} = 5$ GeV/fm$^3$ as given in
Ref.~\cite{BRAHMS-White}, and $\tau_f/\tau_0=8\pm 2$ fm/c, we find
an initial energy density of $\varepsilon_c = (2.0\pm
0.1)\varepsilon_{Bj}=10.0\pm 0.5$ GeV/fm$^3$. If the evolution
deviates from a $1+1$ dimensional perfect flow, then our
estimation is only a lower limit for the initial energy density.

{\it Life-time determination:}
For a Hwa-Bjorken type of accelerationless, coasting  longitudinal flow,
Sinyukov and Makhlin~\cite{Makhlin:1987gm}
determined the longitudinal length of homogeneity as $R_{{\rm long}} = \sqrt{\frac{T_f}{m_t}}\tau_{Bj}$.
Here $m_t$ is the transverse mass and $\tau_{Bj}$ is the (Bjorken) freeze-out time.
However, if the flow is accelerating, the estimated origin of the trajectories is
shifted, so the life-time of the reaction is under-estimated by $\tau_{Bj}$.
(This was pointed out also in refs.~\cite{Wiedemann:1999ev,Csanad:2004cj,Renk:2003gn,Pratt:2006jj}.)
From our solution (c) we obtain 
\bl{Rlong-c}
R_{{\rm long}}=\sqrt{\frac{T_f}{m_t}}\frac{\tau_{{\rm c}}}{\lambda}
    \quad\Rightarrow\quad
    \tau_{{\rm c}} = \lambda \tau_{Bj} .
\ee
BRAHMS data of Fig.~\ref{fig:3} yield
$\lambda=1.18\pm 0.01$, and imply a 18 $\pm$1 \% increase in the estimated $\tau_c$. 

{\it Relation to earlier solutions:}
In our case, similarly to the Hwa-Bjorken case, the initial condition can be given on a $\tau=\tau_0$
hypersurface in the forward light-cone, or on any $\tau_0(\eta)$ continuous Cauchy-surface.
Note, however, that we discuss smooth initial conditions on this 
initial hypersurface,
hence the Landau solution, that starts from a step function, a finite box filled with a constant energy density,
will not be part of the new family of solutions presented below: in our case,
we solve the same dynamical equations as Landau and Khalatnikov, but with modified
boundary conditions. Another similarity to the Landau-Khalatnikov solution is
that in certain limiting cases, we obtain nearly Gaussian rapidity distributions.
Our rapidity distributions are characterized by two parameters, the scale (rapidity density at mid-rapidity) and a shape parameter, which measures the acceleration
effects. In the accelerationless case, the Hwa-Bjorken limit is recovered exactly,
both for the flow profile in the forward light-cone, 
and for the flat rapidity density. We however find external solutions too,
that are valid outside the light-cone. Similarly to the Landau-Khalatnikov solution,
the initial condition outside the light-cone can be specified at $t= 0$, where
the matter is at rest, $v(t=0,r) = 0$. However, in our case, the initial
energy density has an inhomogeneous distribution even in the case of these external solutions, hence $p(t=0,r) $ is never a step function in our family of solutions, 
in contrast to the Landau initial conditions. It is also interesting to mention,
that in the case of our solutions in the future light cone, the acceleration
tends to zero for late times at any given location, but this limit is not uniform.
For example, in case of the $\lambda = 2$ solutions, the acceleration
vanishes for late times at any position, 
but it remains constant along the fluid lines.

{\it In summary}, we have presented a new family of accelerating, exact and simple
solutions of relativistic hydrodynamics.
These new solutions 
are simple, although their finding
was a complicated process that lasted for decades.
BRAHMS pointed out before~\cite{Bearden:2004yx}, that
the rapidity distribution of negative pions
in Au+Au collisions at $\sqrt{s_{NN}} = 200$ GeV
is flatter than a Gaussian, but not completely flat,
hence neither the Landau-Khalatnikov, nor the Hwa-Bjorken solution describes it.
Our new exact solutions 
describe well 
these BRAHMS observations.
We have found  that at least
10$\pm$0.5 GeV/fm$^3$ initial energy densities are reached
at $\tau_0=1$ fm in Au+Au collisions at RHIC.
We have also given an advanced estimate of the life-time of the reaction.
Both estimates include work effects for the first time,
and connect initial conditions and final hadronic observables
with simple and explicit 
formulas. 

As an outlook, the results presented here could be applied to advanced
estimates of initial energy densities in relativistic heavy ion reactions
from CERN SPS through RHIC to LHC. In the limit when the rapidity distribution is flat, the Bjorken energy density estimate is recovered. However, for 
rapidity distribution with a finite width, an advanced formula is found, which yields increased values
of the initial energy density as compared to the Bjorken estimate.

Although we have proven, that our solutions are unique in the considered
general class of parametric solutions of hydrodynamics, 
more work is necessary to investigate the stability and 
possible further generalizations of these solutions. In particular, exact 
solutions are with ellipsoidal symmetry and relativistic acceleration 
are yet to be found, but would be most interesting, as they could provide
new insights to the connection between
elliptic flow data and initial conditions.
Also, exact solutions with more general equations 
of state would be most interesting. These could allow for the investigation of
the dependencies  on the speed of sound of the initial energy
density estimates.

A more detailed and significantly longer description of the results 
summarized above is being prepared and will be submitted for a publication
separately.

\paragraph{Acknowledgements:}
We thank to A. Bialas, T. S. Bir\'o, L. P. Csernai, W. Florkowski, M. Gyulassy, D. Kharzeev,
Yu. Karpenko, I. Mishustin, S. Pratt, Yu. M. Sinyukov, H. St\"ocker, K. Tuchin and W. A. Zajc for inspiring discussions.
T. Cs. thanks Y. Hama and T. Kodama for
inspiring discussions and for  an invitation at an early stage of this study.
This work was supported by the US-Hungarian Fulbright Foundation,
the Hungarian OTKA grants T038406 and T049466,
the FAPESP grants 99/09113-3 00/04422-7 and 02/11344-8 of S\~ao Paulo, Brazil
and the NATO PST.CLG.980086 grant.

\end{document}